\documentclass[twocolumn,10pt,preprintnumbers,amsmath,amssymb]{revtex4}
\usepackage{graphicx}
\usepackage{dcolumn}
\usepackage{bm}
\def\bea {\begin{eqnarray}}
\def\eea {\end{eqnarray}}

\def\be {\begin{equation}}
\def\ee {\end{equation}}

\begin{document}
\title{ The $\eta /s$ of the LQCD-EoS complied hadron gas of different sizes approach common minimum near the crossover temperature}
\author{\it Nachiketa Sarkar}
\author{\it Premomoy Ghosh}
\email{prem@vecc.gov.in}
\address {Variable Energy Cyclotron Centre, HBNI, 1/AF Bidhan Nagar, Kolkata 700 064, India}   
\begin{abstract}
We study the temperature dependence of the ratio of the shear viscosity to entropy density for the LQCD-contrasted hadron resonance gas of different 
finite system-sizes, which may represent the final state hadronic matter, formed in systems of ultra-relativistic collisions. The transport coefficient reaches 
the lowest common value, for the systems of different sizes of thermalized hadron gas, near the critical temperature $T_{c}$ of the QCD crossover.   
\end{abstract}

\maketitle

Following the discovery \cite {ref01, ref02, ref03, ref04} of the Quark-Gluon Plasma (QGP) \cite {ref05, ref06} at the Relativistic Heavy Ion Collider (RHIC) at BNL, in $AuAu$ 
collisions at the centre-of-mass energy ($\sqrt s_{NN}$) of 130 and 200 GeV, the collisions at comparatively lower $\sqrt s_{NN}$ and among  lighter nuclei revealed the 
QGP-like signals, inspiring the RHIC community to take up the Beam-Energy Scan (BES) program aiming for the critical point in the QCD phase diagram. The BES program 
reveals the striking feature \cite {ref07} of collectivity in $AuAu$ collisions at $\sqrt s_{NN}$ = 7.7 GeV, in terms of transverse momentum dependent elliptic flow ($v_{2} (p_{t})$) 
of magnitude similar to that observed in $AuAu$ collisions at $\sqrt s_{NN}$ = 200 GeV. On the other hand, the Large Hadron Collider (LHC) at CERN, while providing collective 
medium of particle production through planned heavy-ion program by creating hotter partonic matter with increased energy density, volume and lifetime \cite {ref08} in collisions 
with heavier nuclei ($PbPb$) at higher $\sqrt s_{NN}$ (2.76 and 5.02 TeV), has also revealed \cite {ref09, ref10, ref11} the unexpected collective behaviour of particle productions 
in high-multiplicity events of $pp$ collisions at $\sqrt s$ = 7 and 13 TeV.  \\

In order to understand the collective behavior of particle production from systems formed in widely varied combinations of colliding particles / nuclei with different centrality and 
the centre of mass energy of collisions, comparative studies in common approaches or models are imperative. The surprisingly large $v_{2}(p_{T})$ at lower RHIC energies 
\cite {ref07} has been addressed \cite {ref12} in terms of the ratio of shear viscosity over the enthalpy density multiplied by the temperature, $\eta T / (\epsilon + p)$, a measure of 
fluidity for a system of hadron gas with a finite baryon chemical potential, $\mu_{B}$. In case of zero chemical potential, this quantity reduces to the ratio of shear viscosity and 
entropy density $(\eta/ s)$. There are several studies \cite {ref13, ref14, ref15, ref16, ref17, ref18, ref19} on the transport properties of hadron resonance gas (HRG) of infinite 
size at zero chemical potential in terms of $\eta/ s$. However, for realistic comparison of systems of final state hadrons in relativistic collisions of proton-proton, proton-nucleus 
and nucleus-nucleus collisions, one should ideally consider HRG of finite system-size.\\

In this article, we estimate and compare the ratio of shear viscosity to entropy density $(\eta/ s)$ for different finite sizes of hadron resonance gas at zero chemical potential 
following molecular kinetic theory approach \cite {ref13, ref14, ref15, ref16, ref17}, relating shear viscosity coefficient to average momentum transfer. We consider hadron resonance 
gas of different finite sizes, which comply with LQCD EoS at zero chemical potential, to be equivalent to final state of particle production from collective medium of different 
sizes formed in different systems of ultra-relativistic collisions at the top RHIC and LHC energies.\\

Recent LQCD calculations \cite {ref20, ref21, ref22, ref23} confirm existence of two phases of strongly interacting matter: the de-confined thermalised partonic matter at temperature
higher than the critical temperature $T_{c}$ and a confined state of hadronic matter below $T_{c}$. The LQCD data \cite {ref24, ref25, ref26, ref27} of hadronic phase at temperature
lower than $T_{c}$ can be well reproduced by Hadron Resonance Gas (HRG) model, usually constituted with hadrons and resonances of known masses as provided in mass tables 
of Particle Data Groups (PDGs). The version of the HRG with Van der Walls excluded volume (EV) approximation \cite {ref28, ref29} introduces repulsive interaction among the
constituents of the hadron gas, enabling one to estimate the transport coefficient, $\eta /$s of the medium. The addition of exponentially increasing continuous hadronic mass 
spectrum, suggested by R. Hagedorn \cite {ref30}, lowers the value of the transport coefficient of the hadron gas at zero chemical potential to match the LQCD data better \cite 
{ref14, ref15} at the QCD critical temperature, $T_{c}$. Several other works demonstrate \cite {ref31, ref32, ref33, ref34} that the Hagedorn mass spectrum contributes significantly 
to the QCD equation of state below the critical temperature ($T_{c}$). It is worth mentioning at this point that the consideration of Hagedorn states over and above the experimentally
identified hadrons and resonance states, listed by the PDGs, is complemented with the recent Lattice QCD works \cite {ref35, ref36} which indicate the need for inclusion of as yet
unobserved resonances to HRG to match the LQCD calculations.\\
 
In the HRG model, the grand canonical partition function of ideal, non-interacting hadron resonance gas is written as \cite{ref27}:\\
\begin{equation}
\ln Z\textsuperscript{id}=\sum_{i=1} \ln Z_{i}^{id}
\end{equation}
where the sum is over all the hadrons.\\

The $i^{th}$ particle partition function is written as:\\
\begin{equation}
\ln Z_{i}^{id}=\pm\frac{V g_i}{2\pi^2 }\int_{0} ^\infty p^2 dp  \ln \left\{1\pm \exp[-(E_i-\mu_i)/T] \right\},
\end{equation}

where $E_{i}=\sqrt{p^2+m^2_{i}}$ and $\mu_{i} = B_i\mu_B + S_i\mu_s + Q_i\mu_Q $. The symbols carry their usual meaning. The $(+)$ and $(-)$ sign corresponds to 
fermions and bosons respectively.\\

The thermodynamical variables, the pressure ($P(T)$), the energy density ($\epsilon (T)$) and the entropy density ($s(T)$) for ideal hadron resonance gas of infinite volume 
at zero chemical potential can be written as :\\
\begin{widetext}
\begin{eqnarray}
P^{id}(T) =\frac {T}{V} \ln Z^{id}=\pm \sum_i\frac{g_iT}{2\pi^2 } \int_{0}  ^\infty p^2 dp  \ln \left\{1\pm \exp[-E_i/T] \right\}
\label{eq:ideal_pressure}\\
\epsilon^{id}(T) =\frac {E^{id}}{V}=-\frac{1}{V} \Big(\frac{\partial  \ln Z^{id}}{\partial \frac{1}{T}}\Big) =\sum_i \frac{g_i}{2\pi^2 }\int_{0}  ^\infty \frac {p^2 dp}  {\exp[E_i/T] \pm 1 }E_i
\label{eq:ideal_energy}\\
s^{id}(T) =\frac{1}{V} \Big(\frac{T( \partial  \ln Z^{id})}{\partial T}\Big)_{V} 
=  \pm\sum_i\frac{g_i}{2\pi^2 }\int_{0}  ^\infty  {p^2 dp} \Big[ \ln \left\{1\pm \exp[-E_i/T] \right\}\pm \frac{E_i}   {  T\big( \exp[E_i/T] \pm 1 \big)}\Big ]
\label{eq:ideal_entropy}
\end{eqnarray}
\end{widetext}
The trace anomaly $I^{id}(T)$ for the considered system can be written as:\\
\begin{equation}
I^{id}(T)=\frac{(\epsilon^{id}-3P^{id})}{T^4}
\label{eq:Tanomaly}
\end{equation}

The effect of repulsive interactions at the short distances is introduced in the HRG through the Van der Waals Excluded Volume (EV) method \cite {ref28} by 
considering hard core radius, $r$ of the constituents of the system. The volume of the system is substituted with an effective volume obtained by excluding the sum 
of volume, $v$ = 16$\pi r^3$/3 of the constituent hadrons. The thermodynamic variables, given by equations~{\ref{eq:ideal_pressure}, \ref{eq:ideal_energy} and 
\ref{eq:ideal_entropy}} with Boltzmann approximation and excluded volume effect become:\\
\begin{equation}
P_{EV}(T,\mu) =\kappa_{EV} P^{id}(T,\mu)\\
\label{eq:ev_pressure}
\end{equation}
\begin{equation}
\epsilon_{EV}(T) = \frac{\kappa_{EV}{\epsilon^{id}(T)}}{1 + {v}{\kappa_{EV}}{n}^{id}(T)}
\label{eq:ev_energy}
\end{equation}
\begin{equation}
s_{EV}(T)=\frac{\kappa_{EV} s^{id}(T)}{1+v\kappa_{EV} n^{id}(T) }
\label{eq:ev_entropy}
\end{equation}
where ${\kappa_{EV}}$ (\textless 1) is the excluded volume suppression factor, given by $\kappa_{EV} = \exp(-{{v}{P_{EV}}} / {T})$. \\

The Hagedorn mass spectrum can be written as \cite{ref37}:
\begin{equation}
\rho(m) = C \frac{\theta(m-M_0)}{(m^2+m^2_0)^a}\exp(\frac{m}{T_H})
\end{equation}
The thermodynamic variables for a system of hadron resonance gas including the Hagedorn states and without the excluded volume effect, can be written as: \\
\begin{widetext}
\begin{eqnarray}
P^{H}(T) =\frac{T}{2\pi^2 }\int  dm   \int_{0}  ^\infty p^2 dp \exp\Big (-\frac{\sqrt{m^2+p^2}}{T}\Big)\Big[\sum_i g_{i}\delta(m-m_{i})+\rho(m)\Big]\\
\epsilon^{H}(T) = \frac{1}{2\pi^2 }\int dm \int_{0}  ^\infty {p^2dp}\sqrt{m^2+p^2}\exp \Big (-\frac{\sqrt{m^2+p^2}}{T}\Big)\Big[\sum_i g_{i}\delta(m-m_{i})+\rho(m)\Big]\\
s^{H}(T) = \frac{1}{2\pi^2 }\int dm \int_{0} ^\infty {p^2 dp} \exp \Big (-\frac{\sqrt{m^2+p^2}}{T}\Big)\Big(1+\frac{\sqrt{m^2+p^2} }{T}\Big)\Big[\sum_i g_{i}\delta(m-m_{i}) +\rho(m)\Big]
\end{eqnarray}
\end{widetext}

The thermodynamic variables, $P^{H}_{EV}(T)$, $\epsilon^{H}_{EV}(T)$ and $s_{EV}^{H}(T)$ for a system of hadrons, resonances and Hagedorn states with excluded volume 
effect at zero chemical potential and with the Boltzmann approximation can be written in the form of equations~{\ref{eq:ev_pressure}, \ref{eq:ev_energy} and \ref{eq:ev_entropy}}, 
where ${\kappa^{H}_{EV}}$, the excluded volume suppression factor for system including Hagedorn states, takes the form, $\kappa^{H}_{EV} = \exp(-{{v}{p^{H}_{EV}}} / {T})$. \\

We aim to study the system-size dependent transport coefficient for the finite-size systems of hadron gas including the Hagedorn states. In literature, there exist different 
HRG-based calculations with different mass-tables of varied ranges of hadrons and resonances. These studies also consider different hard-core radius of constituent hadrons 
for implementing the Excluded volume effect and the parameters related to the Hagedorn mass spectrum. To define the system for our targeted study, we choose, depending 
on previous studies, the values of the system-parameters which are optimized \cite {ref32, ref38} for simultaneous consideration of the EV effect and inclusion of the Hagedorn
States (HS). We set the Hagedorn parameters for our study as: $M_0$ = 2.0 GeV, $T_H$ = 160 MeV, $m_0$ = 0.5 GeV, $a = \frac{5}{4}$ and $C = 0.05 GeV^{3/2}$. All through 
this work, we consider the mass table provided in Ref. \cite {ref39} for the experimentally identified hadrons and resonance states, while the maximum of the Hagedorn-state mass, 
$M_{Max}$ equals 6 GeV. We compare our calculations with the LQCD results provided in ref. \cite {ref23}.\\

\begin{figure}[htb!]
\centering
\includegraphics[scale=0.40]{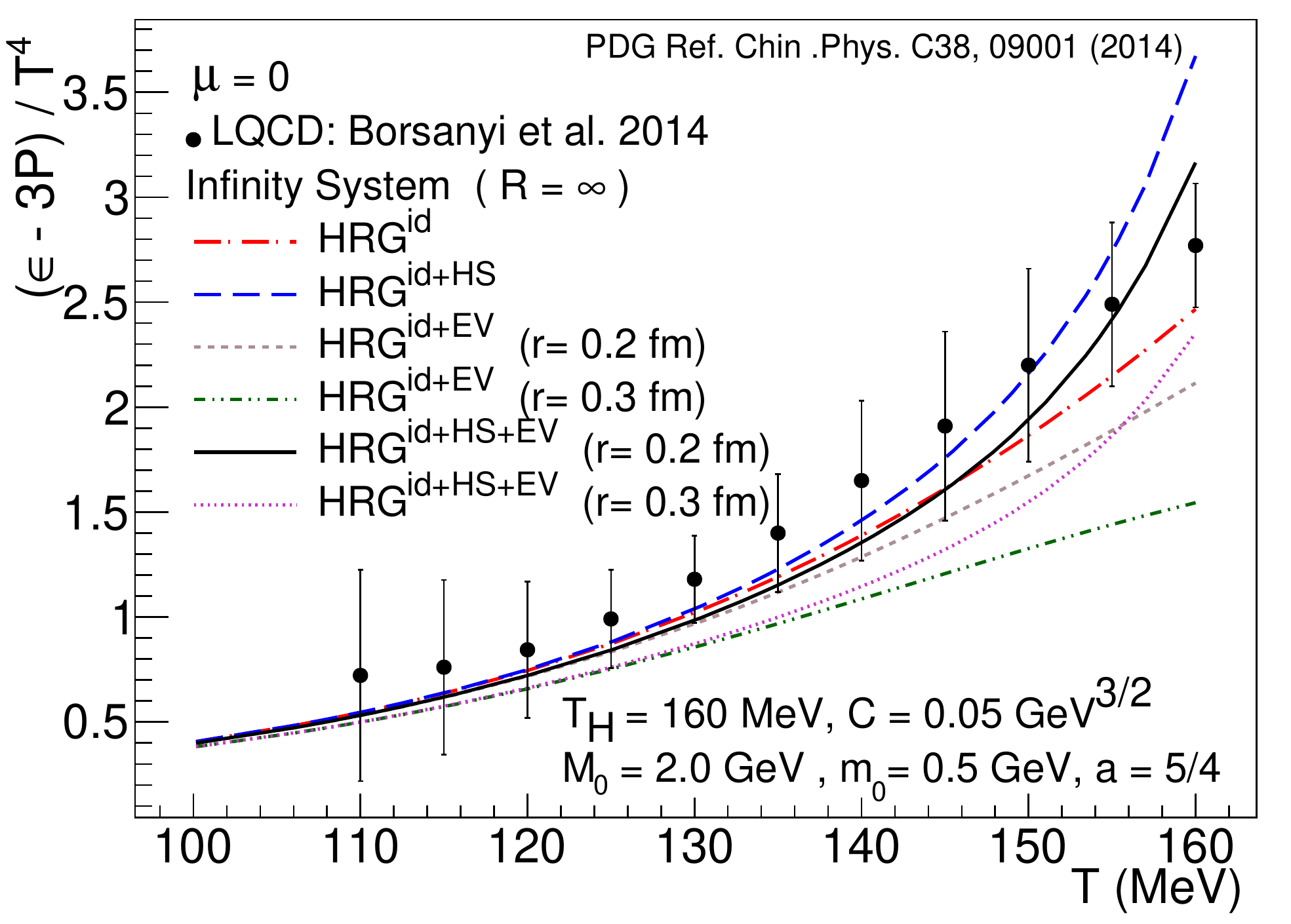}
\caption{The trace anomaly, ${(\epsilon - 3P)}/{T^4}$ as a function of $T$ for different systems, i) ideal HRG, ii) ideal HRG + HS, iii) ideal HRG + EV effect (with the 
hard core radius, $r$ = 0.2 and 0.3 fm), iv) Ideal HRG + HS + EV effect (with $r$ = 0.2 and 0.3 fm), with Boltzmann approximations, is compared with LQCD data 
\cite {ref23}.}
\label{fig:traceanomaly_1} 
\end{figure}
In the Figure~\ref{fig:traceanomaly_1}, we present the results on the trace anomaly, ${(\epsilon - 3P)}/{T^4}$ as a function of $T$, calculated for different systems, 
 i) ideal HRG, ii) ideal HRG + HS, iii) ideal HRG + EV effect, iv) Ideal HRG + HS + EV effect with varied hard-core radius ($r$ = 0.2, 0.3 fm) and Boltzmann approximations. 
The LQCD data \cite {ref23} are included in the figure for comparison. It is clear from the Figure~\ref{fig:traceanomaly_1}, that the ideal HRG can explain the LQCD data 
up to $\textless$ 145 MeV and that for explaining the LQCD data beyond 145 MeV, one indeed needs to include Hagedorn States. Also, for an infinite system of hadron 
gas, the excluded volume effect with hard-core radius $r$ = 0.2 fm explains the LQCD data better than the larger values of $r$. This observation is consistent with the 
study in Ref. \cite {ref32, ref33}.\\
 
\begin{figure}[htb!]
\centering
\includegraphics[scale=0.40]{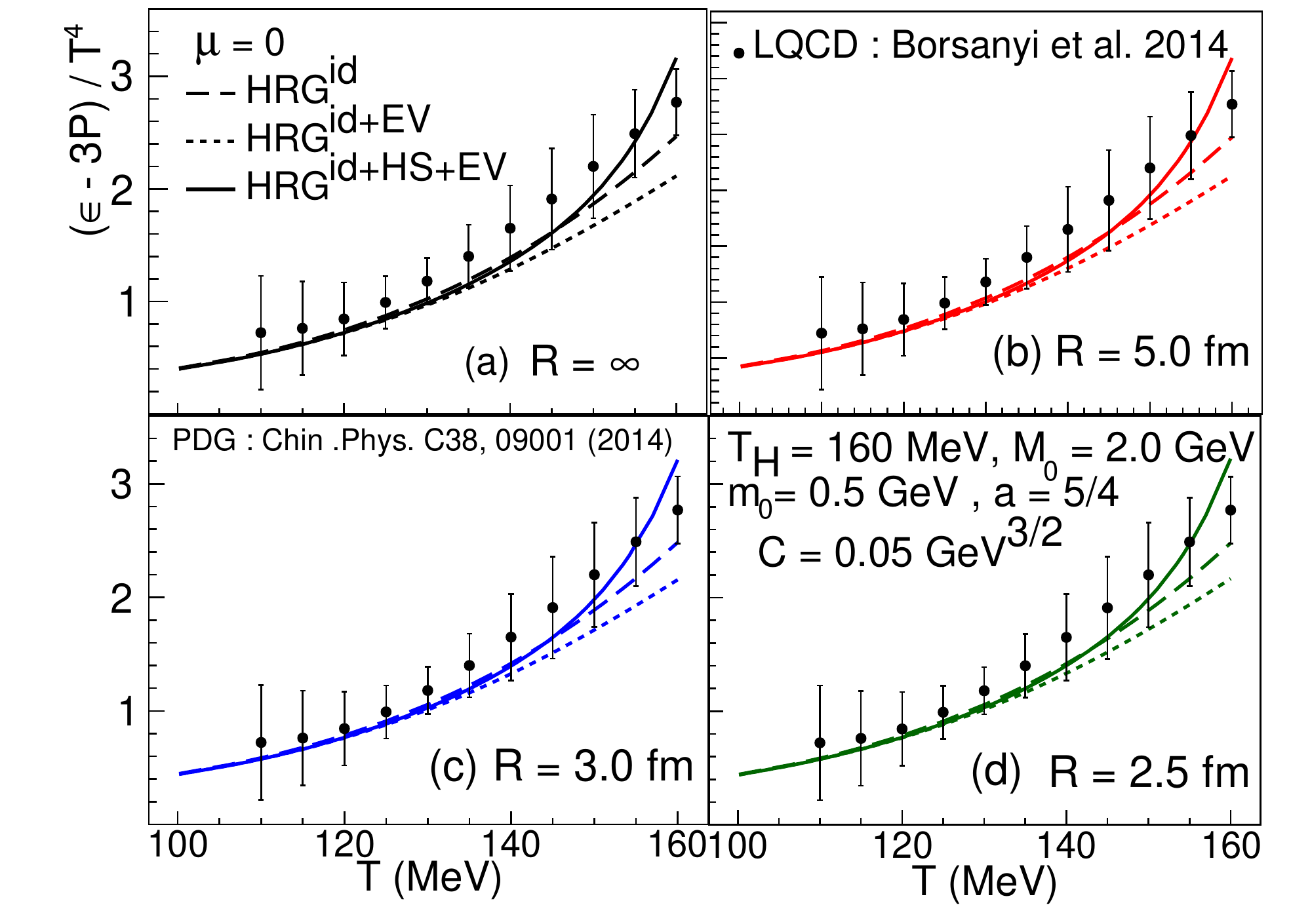}
\caption{The trace anomaly, ${(\epsilon - 3P)}/{T^4}$ as a function of $T$ for different systems, i) ideal HRG, ii) ideal HRG + EV effect, iii) Ideal HRG + HS + EV effect for 
different system size, with Boltzmann approximations, is compared with LQCD data \cite {ref23}.}
\label{fig:traceanomaly} 
\end{figure}

\begin{figure}[htb!]
\centering
\includegraphics[scale=0.40]{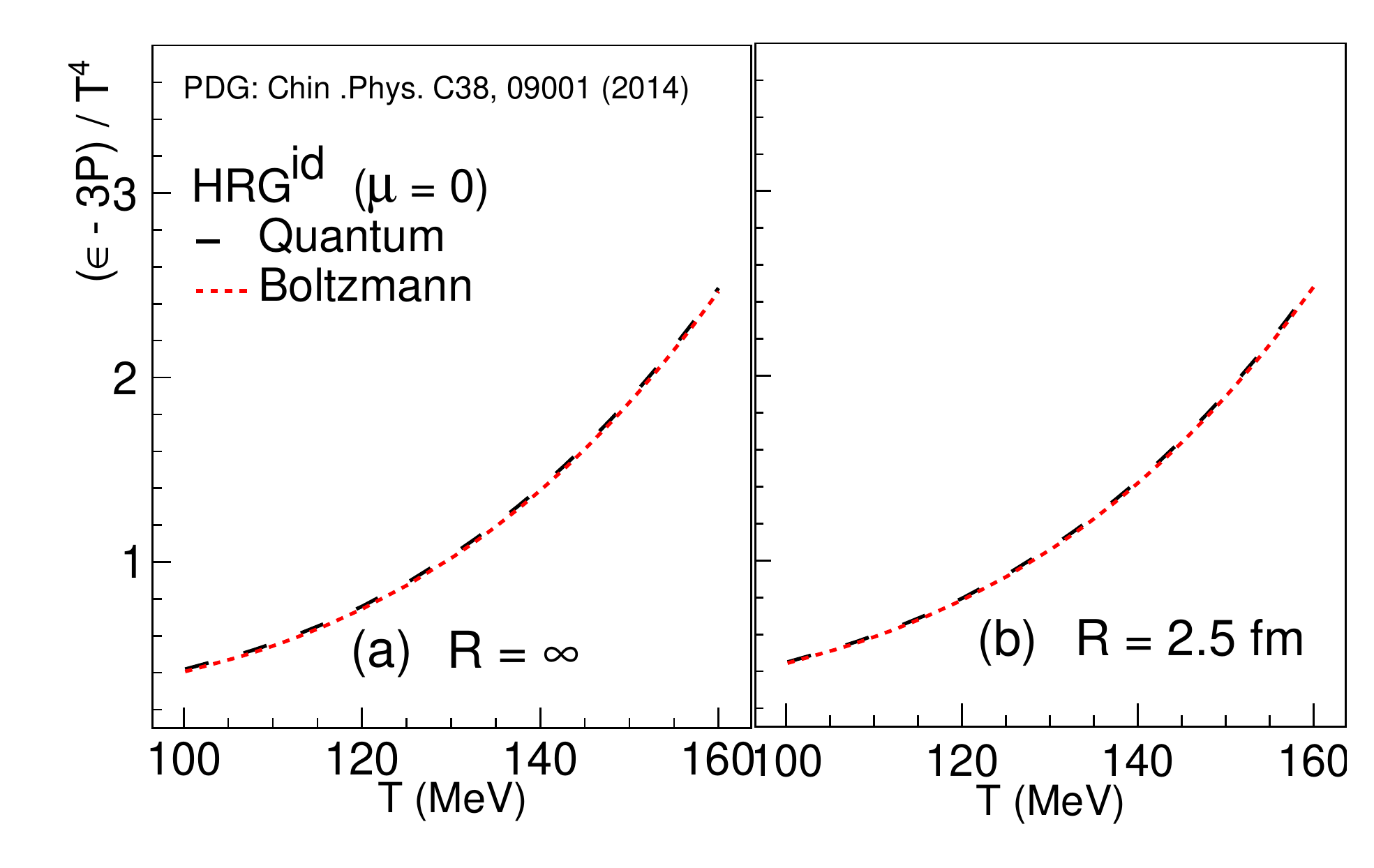}
\caption{The trace anomaly, ${(\epsilon - 3P)}/{T^4}$ as a function of $T$ for different system size of ideal HRG, implemented using the lower limit of momentum, 
with classical and quantum statistics.}
\label{fig:traceanamoly_boltzmann} 
\end{figure}
The finite size effect to thermodynamic system can be implemented by applying low-momentum cuts \cite {ref38, ref40, ref41} in the particle phase-space. A simple proof of 
principle, in this respect, has been presented in ref. \cite {ref40}, where, with a model of one-dimensional gas of noninteracting bosons, it has been shown that the momentum 
cutoff is connected with the system size. The variation in system-size in our study has been incorporated by cutting off different low momentum states ensuring that the temperature 
dependence of thermodynamic variables for these system-sizes lie within the error bars of the lattice calculations. To put it precisely, the momentum cuts are implemented by using 
the lower limit of integral over momentum space in the thermodynamic functions as $p_{cutoff} (MeV)$ = 197 ${\pi} / {{R} (fm)}$, where $R$ is the characteristic system-size that is
studied as a function of $T$ up to the chemical freeze-out. To study the system-size dependence of $\eta /s$ for hadron gas we choose the system-sizes which have already been 
shown \cite {ref38} to be consistent with the LQCD EoS. The HRG system with $R < $ 2.5 fm cannot reproduce LQCD calculations. For $R >$ 5 fm, the effect of the finite sizes reduces
and the system starts behaving more like infinite system.\\
  
\begin{figure}[htb!]
\centering
\includegraphics[scale=0.40]{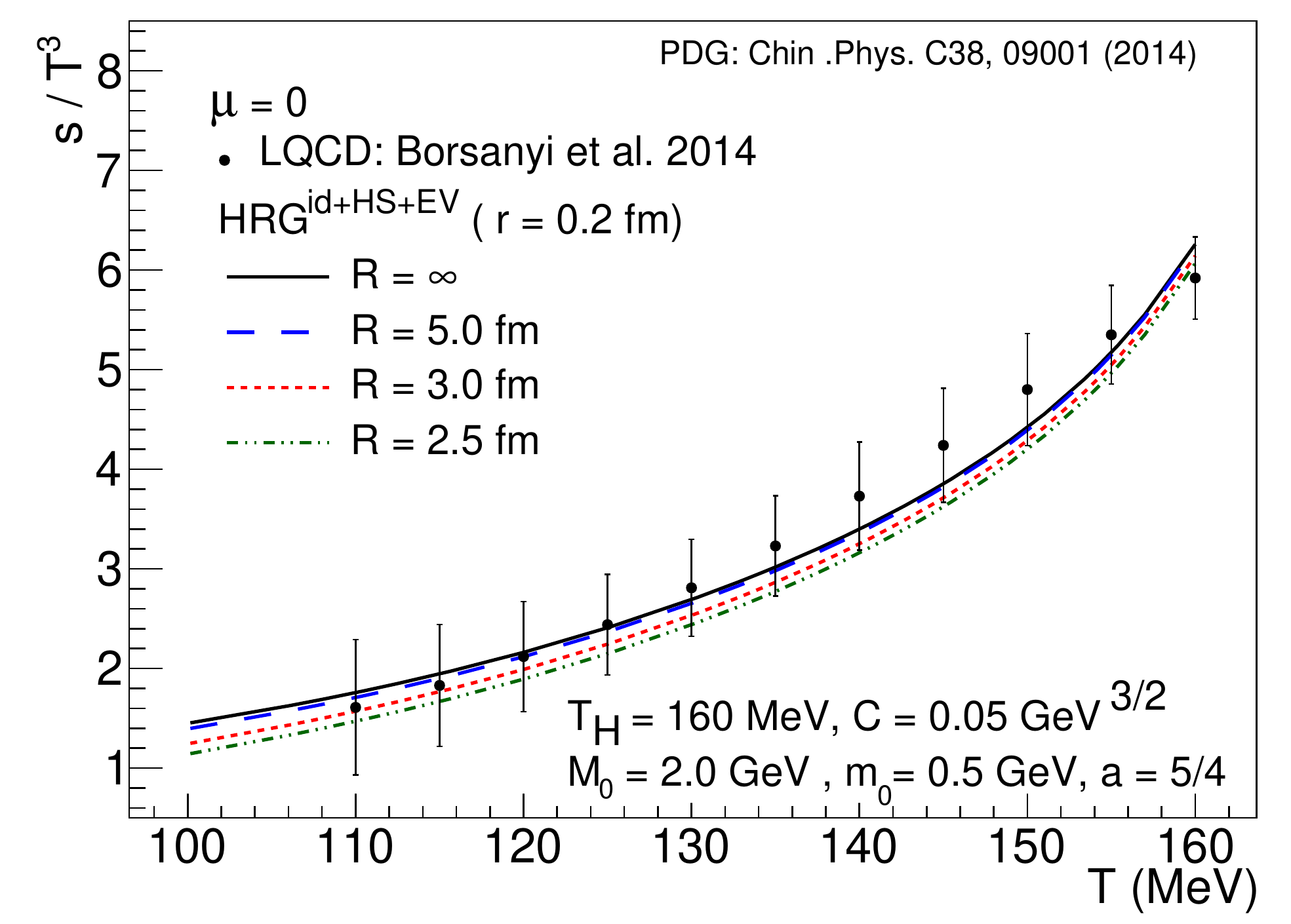}
\caption{The entropy density for different size of system of Ideal HRG + HS + EV effect with $r$ = 0.2 fm as a function of $T$ is compared with LQCD data 
\cite {ref23}.}
\label{fig:entropy} 
\end{figure}

As can be seen in the Figure~\ref{fig:traceanomaly}, the temperature dependent trace anomaly, calculated for hadron gas of different sizes, down to the one - corresponding to 
$R = 2.5$ fm., follow the LQCD calculations. At temperature higher than $\sim 145$ MeV, the agreement with the lattice results is better when the Hagedorn states are added
in the calculations. The inclusion of Hagedorn states implies consideration of Boltzmann approximation. While limiting the size of the system from quantum mechanical 
consideration, it is important to see the effect of Boltzmann approximation for the ideal HRG for different sizes. It is clear from Figure~\ref{fig:traceanamoly_boltzmann} 
that the Boltzmann approximation doesn't really affect thermodynamics of HRG on implementation of finite system-size with momentum-cut.  \\

We calculate the entropy density for hadron gas, including Hagadorn States and the Excluded Volume effect for the considered system sizes. One can see from the 
Figure~\ref{fig:entropy}, the variation in entropy density of different sizes of hadron gas reduces with increasing temperature and entropy density reaches the maximum 
value at $T = $160 MeV, the highest value in the considered temperature range. \\

One can obtain from molecular kinetic theory, the shear viscosity, $\eta$ of an infinite gaseous system proportional to the number density, the mean free path and the average 
momentum of the gas molecules. Accordingly, the shear viscosity ($\eta$) for hadron gas consisting of relativistic hadrons and resonances of discrete mass and of hard 
sphere radius $r$, where $1/r^{2}$ introduces the excluded volume effect, is calculated by analytically developed formula \cite{ref13} as:\\
\begin{equation}
\eta_{EV}=\frac{5}{64\sqrt 8 r^2 }\sum_i \frac{ <|P_i|>n^H_{i(EV)}(T)}{n_{EV}^{H}(T)}
\label{eq:eta_KT1}
\end{equation}

where, $n^H_{i(EV)}(T)=\frac{\kappa^{H}_{EV}n^{id}_i(T)}{1+v \kappa^{H}_{EV} n^{H}(T)} $, $n^{id}_i(T)$ and $n^{H}(T)$ being the number density of ith hadron and of all the hadrons 
(including resonances and Hagedorn states), respectively. Here, ${n_{EV}^{H}(T)}$ is $n^{H}(T)$ with excluded volume correction. $r$ is the hard core radius of hadrons which is considered 
to be the same for all the species. The mean free path  of the system, in terms of the number density and the hard core radius becomes ${1/r^{2}{\sum n^H_{i(EV)}(T)}} = {1/r^{2} {n_{EV}^
{H}(T)}}$. The average momentum ($<|P_i|>$) is given by:\\

\begin{equation}
<|P_i|> = \frac{\int_{0}^{\infty} p^3dp \exp [-\sqrt{(p^2+m_i^2)}/T]}{\int_{0}^{\infty}p^2dp \exp [-\sqrt{(p^2+m_i^2)}/T]}
\end{equation}\\	

\begin{figure}[htb!]
\centering
\includegraphics[scale=0.40]{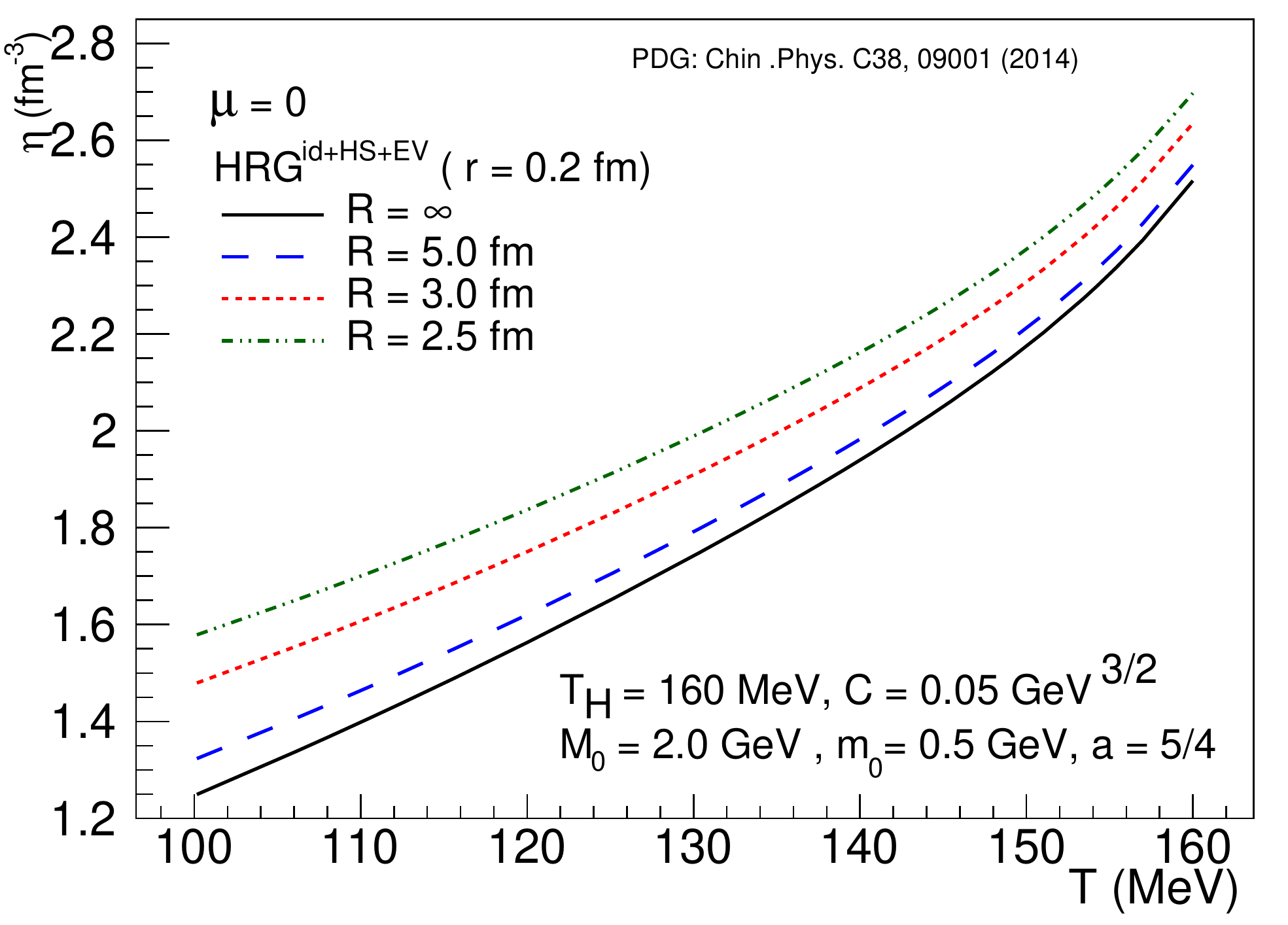}
\caption{The shear viscosity $\eta$ for different sizes of systems as mentioned in the figure caption of Figure~\ref{fig:entropy}.}
\label{fig:eta} 
\end{figure} 

As has been pointed out in ref.~\cite{ref15}, the behavior of the massive Hagedorn states in a hot hadron gas is not yet well studied to calculate the effect of these rapidly decaying 
resonance states on the transport coefficients of the medium, though there have been different methods \cite {ref13, ref14} available in the literature for calculating the transport 
coefficient for the Hagedorn states. Nevertheless, the estimate of the shear viscosity for the Hagedorn States, as suggested in ref.~\cite{ref13}, has been found to be in good 
agreement \cite{ref15} with estimates from other methods. \\

The shear viscosity for continuous Hagedorn mass spectra, is given \cite{ref15} by equation~\ref{eq:eta_HS2}:\\
\begin{equation}
\eta_{EV}^{H} =\frac{5}{64\sqrt{8}r^2}\frac{1}{2\pi^2n^{H}(T) }\int \rho(m) dm\int_{0}  ^\infty p^3 dp \exp \Big (\frac{-\epsilon}{T}\Big)       
\label{eq:eta_HS2}
\end{equation}
So, for a system of hadron gas with resonance states, including the Hagedorn States, the shear viscosity can be written as: \\
\begin{equation}
\eta = \eta_{EV} + \eta_{EV}^{H}
\label{eq:eta_HS3}
\end{equation}

Because of the higher mass, the particle density of Hagedorn states is much smaller than that of the discrete states. So, the contributions to the mean-free path due to interactions 
between the discrete states and the Hagedorn states can be ignored \cite {ref14} for the sake of simplicity in calculations. \\

\begin{figure}[htb!]
\centering
\includegraphics[scale=0.40]{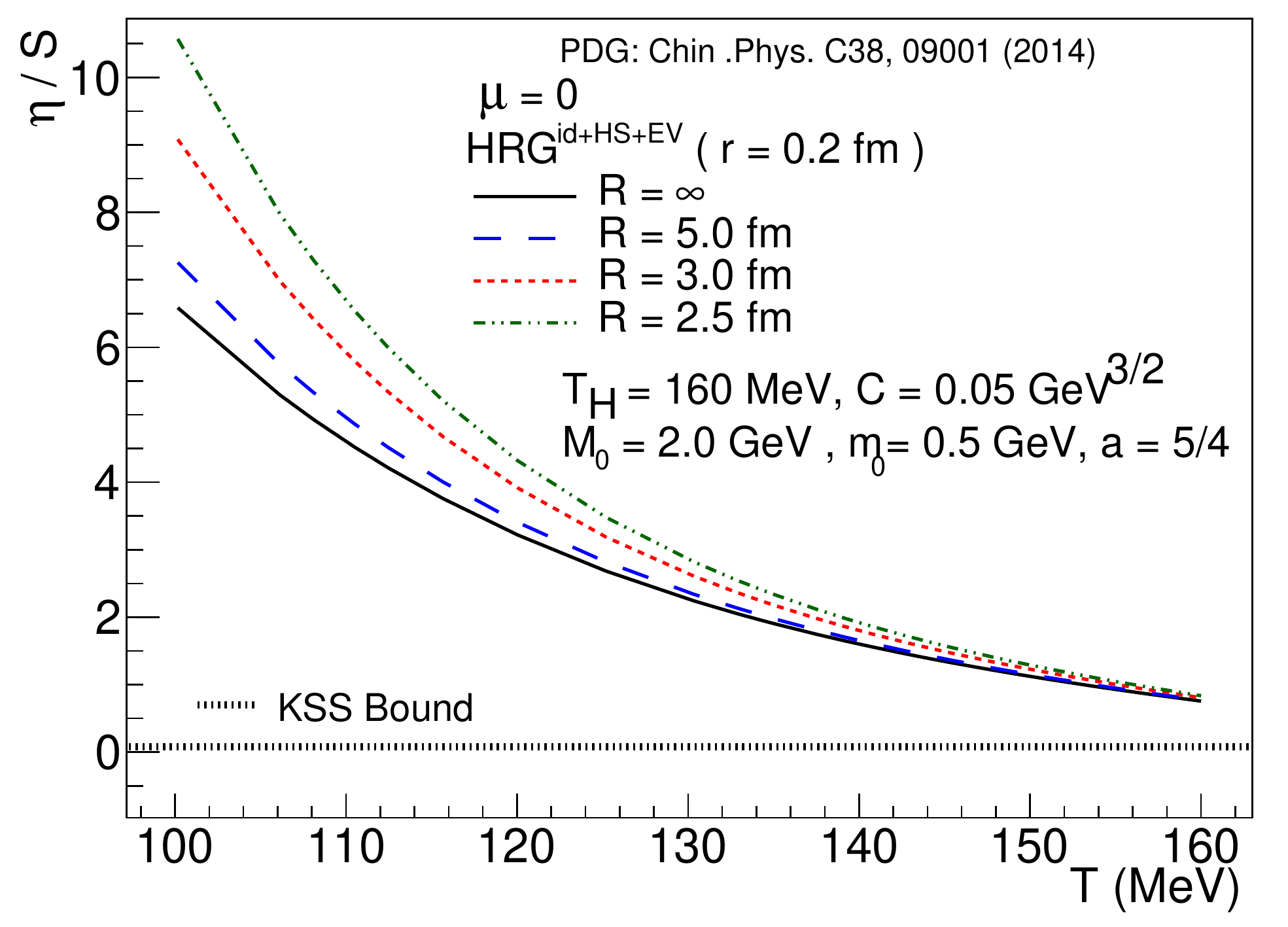}
\caption{The ratio, $\eta /s$ for different sizes of systems as mentioned in the figure caption of figure~\ref{fig:entropy}. The KSS bound \cite {ref42} has been included for reference.}
\label{fig:etabys} 
\end{figure}

We calculate the shear viscosity for the considered finite sizes of hadron gas by implementing corresponding lower momentum cuts and plot in Figure~\ref{fig:eta}, as a function 
of temperature. Like the temperature dependence of the entropy, the shear viscosity also increases with temperature and the variation in the value of shear viscosity for different 
sizes of hadron gas decreases with temperature. The finding of the targeted study of temperature dependence of $\eta/s$ for different sizes of hadron gas, following the LQCD-EoS, 
is depicted in figure~\ref{fig:etabys}, where the conjectured universal lower bound \cite {ref42} of $\eta/ s$ has been shown. In this work, the transport coefficient for infinite hadron 
gas with Hagedorn States and hard-core radius of 0.2 fm., as calculated in \cite {ref15} in the similar approach, has been reproduced to further the study to see the finite-size effect 
of system of hadron gas. We find that at  T = 160 MeV, the temperature close to the QCD crossover \cite{ref21, ref43}, the ratio, $\eta/s$ for different sizes of thermalized hadron 
gas reaches similar value (0.75 for $R = 2.5 fm $ to 0.83 for $R = \infty$). It is worth mentioning that similar nature of $T$-dependence of $\eta /s$ for the considered system-sizes 
is observed also for hadron resonance gas without considering the Hagedorn states. The $\eta/s$ of different sizes of HRG, without Hagedorn states, also tend to meet at $T$ = 160
MeV, though at higher range of values (0.82 for $R = 2.5 fm$ to 0.92 for $R = \infty$). This observation is consistent with previous studies \cite {ref14, ref18} with HRG of infinite size 
at $\mu_{B}$ = 0, where $\eta/s$ reduces with the inclusion of Hagedorn states. We prefer to focus on results of our analysis including the Hagedorn states, as the inclusion of 
these high-mass states are found to describe better the LQCD calculations of temperature-dependent profile of thermodynamic variables of hadron gas system. Further, as Lattice
calculations form our frame of reference, we study the system-size effect of HRG upto the T = 160 MeV, the $T_{C}$ from lattice calculations, and accordingly we chose the 
Hagedorn temperature, $T_{H}$ = 160 MeV. We restrict our calculations up to $T$ = 160 MeV, as at temperature beyond $T_{H}$, the partition function for exponentially increasing 
Hagedorn states diverges \cite {ref44}. However, by extending our study at temperature beyond $T$ = 160 MeV for hadron gas without considering the Hagedorn states we find 
similar tendency of convergence of $\eta /s$ for the considered sizes of the system. One can, thus, conclude that, at low temperature, the system-size of hadron gas affects $\eta /s$ 
and the effect of finite system-size goes off at high temperature.\\

As the finite system-size of the HRG is implemented by cutting-off the low-lying momentum states, at a given low temperature, with higher range of cut on low-lying momentum 
states for a smaller system, the average momentum of hadrons increases, resulting larger $\eta$. The momentum-cut, on the other hand, reduces the number density of the system,
decreasing the $s/T^3$ for smaller system. So, at low temperature, the $\eta /s$ also is more for a smaller system. At high temperature, the effect of cutting-off the low-lying momentum
states on the average momentum diminishes, and the effect of increased particle density becomes predominant factor in determining the transport coefficient. At sufficiently high temperature, 
the increased particle density and consequently the decreased mean free path even for a finite, small system may approach the thermodynamic limit, by appearing infinite in size in comparison 
to mean free path, where $\eta /s$ becomes independent of system-size. For the hadron gas of considered finite sizes, complying the LQCD-Eos, we find the $T$-dependence of $\eta /s$, 
close to the $T_{c}$, tend to reach the value for the infinite system of the hadron gas. \\ 

It is important to note that, in this study of the temperature dependence of the finite-size effect on transport coefficient of hadron resonance gas, all the constituents are considered to 
be hard sphere of same radius, effectively assuming same cross-sections for all the particles irrespective of species and temperature. This assumption, ignoring energy and flavour 
dependent cross-sections, likely to affect the calculation of the absolute value of $\eta$ and so cannot be compared with other calculations, on the same footings, where more 
realistic cross-section values have been taken into account. However, a comparison with another similar analysis in ref.~\cite{ref17}, provides a consistency check. In ref.~\cite{ref17}, 
$\eta /s$ for infinite size of HRG with Hagedorn states reaches 0.7 at $T$ = 170 MeV and $\mu_{B}$ = 0, while the hard core radius for mesons and baryons are taken as 0.4 fm and 0.5 fm, 
respectively. The $\eta /s$ of hadron gas, in the discussed formalism, decreases with increasing hard core radius of constituent particles \cite {ref13}. \\

Although the objective of this study has been to see the finite-size effect on transport coefficient, $\eta /s$ and not the evaluation or comparison of several approaches of the 
measurement, while discussing the $\eta /s$ of hadron gas at $\mu_{B}$ = 0, it may be worth mentioning its values as obtained in different formalisms, involving elaborate 
theoretical calculations including realistic cross-sections. In ref.~\cite{ref18}, the $\eta/s$ for equilibrated hadronic matter, including Hagedorn states, calculated with Kubo formula 
\cite {ref45}, reaches $\sim$ 0.69 at $T$ = 160 MeV. A microscopic transport calculations \cite {ref19} with UrQMD for hadronic phase (without considering Hagedorn states) at 
$\mu$ = 0, employing Kubo formalism, results in a minimum $\eta/s \sim $ 0.9. In literature, there exist many other studies on $\eta /s$ for pion gas or for mixture of pions and 
nucleons with a wide variation in estimation. All these indicate that theoretical formalism for calculating transport coefficient for a multicomponent hadron gas is still developing. 
Also, there is no theoretical attempt yet to include finite-size effect on shear viscosity of hadron gas. At this stage, to study the finite size effect on $\eta /s$ of hadron resonance 
gas, in this phenomenological study, we apply lower limit of the integral over momentum space in calculating both, the thermodynamic variables and the transport coefficient.\\

In summary, we present our study on temperature dependence of transport coefficient, $\eta/s$ for system of hadrons, resonances and Hagedorn states of different finite 
system-sizes, at zero chemical potential. While the $\eta/s$ for different sizes of LQCD-EoS complied hadron gas are different at low temperature, it approaches a common value at 
the QCD crossover temperature. The system-size independence of $\eta /s$ at high-temperature is consistent with the observations reported in references~\cite{ref38} and \cite{ref41}, 
where similar system-size dependent study of hadron gas revealed that the thermodynamic properties of the system do not get affected at high temperature on implementation of 
the finite system-size. Our estimated value of $\eta /s$ at $T$ = 160 MeV for hadron gas at $\mu_{B}$ = 0, with several approximations, particularly ignoring the energy and species 
dependent cross-sections, though not exactly comparable, is not very far from the values obtained from elaborate theoretical calculations, without the approximations. The effect of 
the inclusion of Hagedorn states on the value of $\eta /s$ at the crossover temperature is similar to that observed in those theoretical studies \cite {ref17, ref18}. The approximations 
in our calculations should not affect a comparative study of temperature dependence of $\eta /s$ of different sizes of finite hadron gas and the finding that the tendency of $\eta /s$ of 
the hadron gas of the considered system-sizes becoming size-independent at high temperature arises due to high number density of the system. The observation doesn't change on 
exclusion of Hagedorn states and on extension of the $T$-dependent study beyond $T$ = 160 MeV. From this first ever study of system-size dependence of the transport coefficient 
of hadron gas, considered to be representing final state hadrons produced from thermalised medium of different sizes formed in different systems of relativistic collisions at RHIC and 
LHC (corresponding to $\mu_{B}$ = 0), we infer that as long as the final state hadrons of different system-sizes can be described by LQCD equation of states, the property of the medium 
of hadrons at the QCD crossover can be considered to be independent of the size of the system.   \\

\end{document}